%% file: 0_main.tex
\documentclass[research]{fcs}

\usepackage{bm}
\usepackage{times}
\usepackage{soul}
\usepackage{url}
\usepackage[utf8]{inputenc}
\usepackage{subfigure}
\usepackage{graphicx}
\usepackage{amsmath}
\usepackage{amsthm}
\usepackage{amssymb}
\usepackage{booktabs}
\usepackage{algorithm}
\usepackage{algorithmic}
\usepackage{multirow}
\usepackage[normalem]{ulem}
\useunder{\uline}{\ul}{}
\usepackage{enumitem}
\usepackage{enumerate}
\usepackage{color}
\usepackage{microtype}

\newcommand\norm[1]{\left\lVert#1\right\rVert}

\newcommand{\Set}[1]{\mathcal{#1}}

\newcommand{\ie}{\emph{i.e., }}
\newcommand{\eg}{\emph{e.g., }}

\newcommand{\wrt}{\emph{w.r.t. }}

\title{Graph Convolution Machine for Context-aware Recommender System}
\author[1]{Jiancan Wu}
\author*[1]{Xiangnan He}
\author[2]{Xiang Wang}
\author[3]{Qifan Wang}
\author[1]{Weijian Chen}
\author[4]{Jianxun Lian}
\author[4]{Xing Xie}
\address[1]{School of Information Science and Technology, University of Science and Technology of China, Hefei 230026, China}
\address[2]{5 Prince George's Park, National University of Singapore, Singapore 118404, Singapore}
\address[3]{Google Research, Mountain View, CA 94043, USA}
\address[4]{Microsoft Research Asia, Beijing 100190, China}
\fcssetup{
  received       = {month dd, yyyy},
  accepted       = {month dd, yyyy},
  corr-email     = {xiangnanhe@gmail.com},
}
\begin{abstract}
The latest advance in recommendation shows that better user and item representations can be learned via performing graph convolutions on the user-item interaction graph. However, such finding is mostly restricted to the collaborative filtering (CF) scenario, where the interaction contexts are not available. In this work, we extend the advantages of graph convolutions to context-aware recommender system (CARS, which represents a generic type of models that can handle various side information). We propose \textit{Graph Convolution Machine} (GCM), an end-to-end framework that consists of three components: an encoder, graph convolution (GC) layers, and a decoder.
The encoder projects users, items, and contexts into embedding vectors, which are passed to the GC layers that refine user and item embeddings with context-aware graph convolutions on the user-item graph. The decoder digests the refined embeddings to output the prediction score by considering the interactions among user, item, and context embeddings. We conduct experiments on three real-world datasets from Yelp and Amazon, validating the effectiveness of GCM and the benefits of performing graph convolutions for CARS.
Our implementations are available at \url{https://github.com/wujcan/GCM}.
\end{abstract}

\begin{document}
\input{1_introduction}
\input{2_related}
\input{3_probDef}
\input{4_gcm}
\input{5_experiment}
\input{6_conclusions}

\bibliographystyle{fcs}
\bibliography{fcs}

\begin{biography}{photos/Jiancan_30x33}
Jiancan Wu received his B.S. degree in Electronic Engineering and Information Science from the University of Science and Technology of China (USTC) in 2017. He is currently a Ph.D. student at USTC. His research interests focus on Recommender Systems, Machine Learning and Graph Neural Networks.
\end{biography}

\begin{biography}{photos/xiangnan_30x33}
Dr. Xiangnan He is a professor at the University of Science and Technology of China (USTC). He received his Ph.D. in Computer Science from the National University of Singapore (NUS) in 2016. His research interests span information retrieval, data mining, and multi-media analytics. He has over 70 publications that appeared in several top conferences such as SIGIR, WWW, and MM, and journals including TKDE, TOIS, and TMM. His work on recommender systems has received the Best Paper Award Honorable Mention in WWW 2018 and ACM SIGIR 2016. Moreover, he has served as the PC chair of CCIS 2019, area chair of MM (2019, 2020) ECML-PKDD 2020, and PC member for several top conferences including SIGIR, WWW, KDD, WSDM etc., and the regular reviewer for journals including TKDE, TOIS, TMM, etc.
\end{biography}

\begin{biography}{photos/xiangwang_30x33}
Xiang Wang is now a research fellow at National University of Singapore. He received his Ph.D. degree from National University of Singapore in 2019. His research interests include recommender systems, graph learning, and deep learning techniques. He has published some academic papers on international conferences such as KDD, WWW, SIGIR, and AAAI. He serves as a program committee member for several top conferences such as SIGIR and WWW.
\end{biography}

\begin{biography}{photos/Qifan_30x33}
Qifan Wang received the BS and MS degrees from Tsinghua University, and the PhD degree from Purdue University, all in computer science. He is a researcher in Google Research, and his research interests include machine learning, information retrieval, data mining, computer vision and natural language processing.
\end{biography}

\begin{biography}{photos/weijian_30x33}
Weijian Chen is currently a Ph.D. student at the University of Science and Technology of China(USTC). His research interests focus on User Profiling, Recommender System, and Graph Neural Networks. He has served as a research intern in the JD Data Science Laboratory and a project intern in the Kwai Multimedia Understanding Department.
\end{biography}

\begin{biography}{photos/Jianxun_30x33}
Jianxun Lian is now a senior researcher at Microsoft Research Asia. He received his Ph.D. degree from University of Science and Technology of China in 2018.   His research interests include recommender systems and deep learning techniques. He has published some academic papers on international conferences such as KDD, IJCAI, WWW, SIGIR and CIKM.  He serves as a program committee member for several top conferences such as AAAI, WWW and IJCAI.
\end{biography}

\begin{biography}{photos/Xie_30x33}
Dr. Xing Xie is currently a senior principal research manager at Microsoft Research Asia, and a guest Ph.D. advisor at the University of Science and Technology of China. He received his B.S. and Ph.D. degrees in Computer Science from the University of Science and Technology of China in 1996 and 2001, respectively. He joined Microsoft Research Asia in July 2001, working on data mining, social computing and ubiquitous computing. During the past years, he has published over 300 referred journal and conference papers, won the 10-year impact award in ACM SIGSPATIAL 2019, the best student paper award in KDD 2016, and the best paper awards in ICDM 2013 and UIC 2010. 
\end{biography}

\end{document}

%% file: 1_introduction.tex
\section{Introduction}
Recommendation has become a pervasive service in today's Web, serving as an important tool to alleviate information overload and improve user experience. 
The key data source for building a recommendation service is user-item interactions, \eg clicks and purchases, which spawn wide research efforts on collaborative filtering (CF)~\cite{1,2,3} that leverage the interaction data only to predict user preference. 
Recently, inspired by the success of graph neural networks (GNNs)~\cite{4,5}, researchers have attempted to employ GNNs on recommendation in which CF signals are exhibited as high-order connectivity~\cite{3,6,7,8}.
While CF provides a universal solution for recommendation, it falls short in utilizing the side information of interaction contexts. In many scenarios, the current contexts could have a strong impact on user choice. For example, in restaurant recommendation, the current time and location can effectively filter out unsuitable candidates; in E-commerce, the click behaviors in recent sessions provide strong signal on user next purchase. As such, it is important to develop context-aware recommender system (CARS) that can effectively integrate contexts (and possibly other side information like user profiles and item attributes) into user preference prediction~\cite{9}. 

Inspired by the matrix completion view of CF, early research naturally extended the problem of CARS to tensor completion~\cite{10}, which however suffers from high complexity. 
Later on, Rendle proposed factorization machine (FM)~\cite{11}, which to the first time addressed CARS from the view of standard supervised learning. Specifically, it converts all information related to an interaction to a feature vector via multi-hot encoding, modeling the second-order feature interactions to predict the interaction label.  
Due to its generality and effectiveness, FM soon becomes a prevalent solution for CARS and is followed by many work. For example, in the era of deep learning, Wide\&Deep~\cite{12} and Deep Crossing~\cite{13} replaced the second-order interaction modeling with a neural network for implicit interaction modeling; recently, Neural FM~\cite{14}, Attentional FM~\cite{15}, xDeepFM~\cite{16}, and Convolutional FM~\cite{17} extended FM with various kinds of neural networks to enhance its expressiveness.

Summarizing existing CARS models, we can find a common drawback:   
they follow the standard supervised learning scheme that ignores the relationship among data instances. 
This may limit the model's effectiveness in capturing the CF effect, since it needs to consider multiple interactions simultaneously to recognize the CF patterns. 
An evidence is from the neural graph collaborative filtering (NGCF) work~\cite{3}, which demonstrates that connecting the interactions in the predictive model significantly improves the embedding quality for CF. 
Since in CARS user-item interactions still play an important role by reflecting user preference, it is reasonable to believe that properly modeling the relationship among interactions can improve the model quality. 
Moreover, the recent neural network-based methods like xDeepFM~\cite{16} and Convolutional FM~\cite{17} suffer from low efficiency in online serving, since each candidate item needs be scored separately with the deep model architecture that models complex feature interactions, which could be very time-consuming. 

In this work, we aim to propose new CARS model by addressing the above-mentioned limitations. 
Firstly, we cast the data in CARS as an attributed user-item graph, where the side information of users and items are represented as node features, and the contexts are represented as edge features (Figure \ref{fig:dataProcessing}). Secondly, we propose an end-to-end model that consists of three components: an encoder, graph convolution (GC) layers, and a decoder (Figure \ref{fig:framework}). The encoder projects users, items, and contexts into embedding vectors; the GC layers then exploit the interactions to refine the embeddings via performing graph convolutions; lastly, the decoder models the interactions among embeddings via FM to output the prediction score. 
After the model is trained, the refined embeddings by GC layers can be pre-computed before serving. As such, the time complexity of online serving is the same as FM, being much more efficient than the recent neural network methods. 
We summarize the contributions of this work as follows:
\begin{itemize}
\item We highlight the limitation of the mainstream supervised learning schemes and the necessity of exploiting the relationship among data instances in the predictive model of CARS. 
\item We propose a new model named Graph Convolution Machine (GCM), unifying the strengths of graph convolution network and factorization machine for CARS. 
\item We conduct extensive experiments on three real-world datasets which demonstrate the effectiveness and efficiency of GCM. 
\end{itemize}

%% file: 2_related.tex
\section{Related Work}
\subsection{Context-aware Recommendation}
Extensive studies on context-aware recommender system (CARS)~\cite{11,14,16} have been conducted and achieved great success.
Learning informative representations, based on user-item interactions (\eg clicks, purchases) and contextual features (\eg location, time, last purchase), has been a central theme of research on CARS.
Towards this end, modeling interactions among different features is showing promise.
Early, factorization machine (FM)~\cite{11} embeds each feature into a vector representation, and utilizes inner product to capture their pairwise relationships (\eg the second-order feature interactions).
Due to its generality and effectiveness, FM becomes a prevalent solution for CARS.
Many works resort to this paradigm, such as FFM~\cite{18}.
Recent works~\cite{12,14,15,16,19} leverage deep neural networks to model higher-order feature interactions, so as to generate better representations and enhance recommendation performance. For example,
NFM~\cite{14} proposes a bilinear interaction operation which uses a sum pooling over the pair-wise dot-product of feature vectors;
AFM~\cite{15} learns the importance of each feature interaction via the attention mechanism; 
xDeepFM~\cite{16} extends the Cross Network~\cite{20} to the Compressed Interaction Network (CIN) which models high-order interactions explicitly at vector-wise level; 
while Convolutional FM~\cite{17} models second-order interaction with outer product, forming an interaction cube, then applying 3D convolution to learn high-order interactions.
It's worth mentioning that another research line close to CARS is the CTR (Click Through Rate) prediction~\cite{21,22,23,24}, which also focuses on modeling the complex feature interactions. The key difference lies in the evaluation protocol: most CARS models adopt top-k recommendation protocols, while CTR prediction models measure log loss or AUC metrics on positive/negative samples.

Despite effectiveness, we argue that present works treat user interactions as isolated data instances, while forgoing their relationships (\eg user behaviors happened at the same time and location are highly likely to reflect user preferences).
This would easily lead to suboptimal representations and limit the performance.
We hence aim to explore relationships among user behaviors in this work.

\subsection{Graph Neural Networks for Recommendation}
Another relevant research line is to leverage graph neural networks (GNNs) for recommendation.
In particular, GNN models~\cite{4,5,25} exploit graph structure to guide the representation learning.
The basic idea is the embedding propagation mechanism, which aggregates the embeddings of neighbors to update the target node’s embedding. By recursively performing such propagations, the information from multi-hop neighbors is encoded into the representation of target node.
GNN models has been widely used in many fundamental tasks due to their strong representation ability, spanning from node classification~\cite{26}, link prediction~\cite{27}, to graph classification~\cite{28}, and achieved remarkable improvements.

Inspired by their success, researchers have attempts to employ GNNs on recommendation.
Recent works on collaborative filtering (CF), such as NGCF~\cite{3}, GC-MC~\cite{29}, SpectralCF~\cite{8} and PinSage~\cite{30}, reorganize historical user behaviors in the form of a user-item bipartite graph, exhibit CF signals as high-order connectivity, and encode such signals into representations.
For CTR prediction task, Fi-GNN~\cite{31} takes multi-field features into consideration by constructing feature graph for each instance and converting the task of modeling feature interactions among fields into modeling node interactions on the feature graph;
GIN~\cite{32} models implicit user intention by the multi-layered intention diffusion and aggregation on the co-occurrence click relationship graph;
\cite{33} builds the multi-relational item graph and applies GNN to capture complex transition relations between items in user bahavior sequences.
Moreover, GNN models have also been employed on other recommendation tasks, including social recommendation~\cite{17,34}, sequential recommendation~\cite{35,36}, and knowledge-aware recommendation~\cite{37,38}.
As such, aggregating useful information from multi-hop neighbors is able to achieve better expressiveness, than single ID embeddings.
Hence, it is reasonable to believe that graph learning is a promising solution to properly model the relationships among interactions.

%% file: 3_probDef.tex
\section{Problem Definition}
\begin{figure}
\centering
\includegraphics[width=0.48\textwidth]{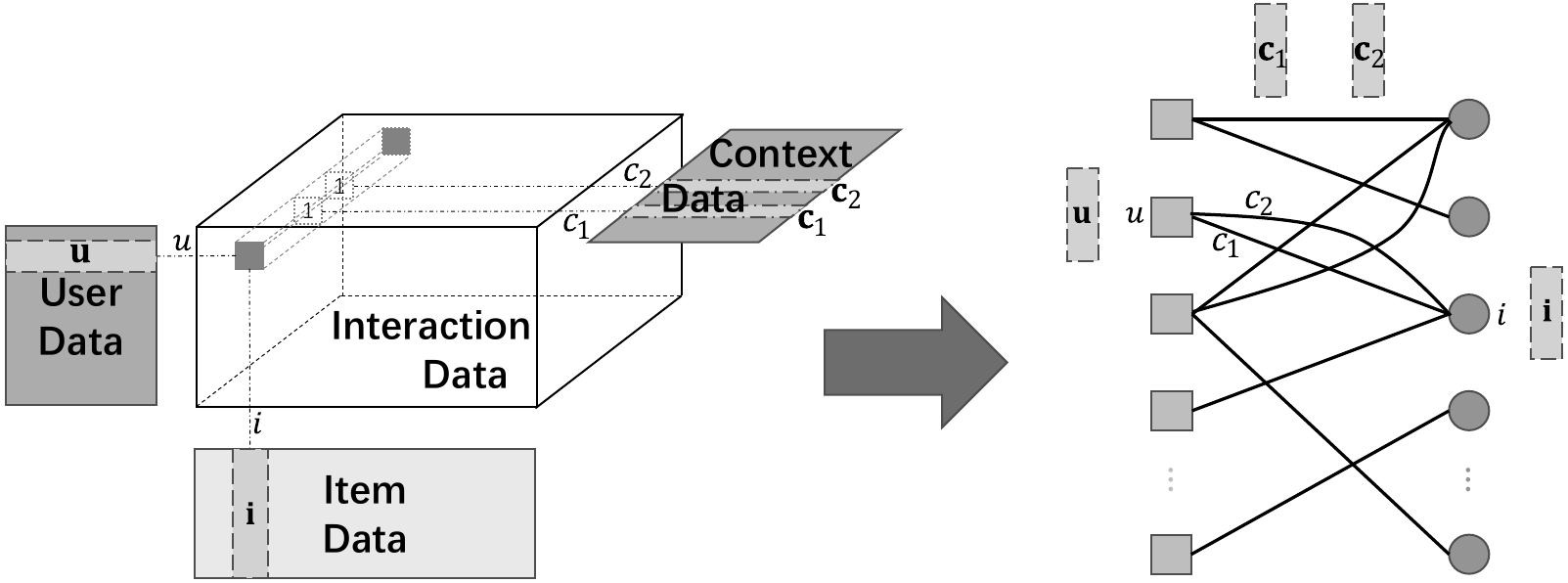}
\caption{The data used for building a CARS. The mixture data of interaction tensor and user/item/context feature matrices are converted to an attributed user-item bipartite graph without loss of fidelity.}
\label{fig:dataProcessing}
\end{figure}

We divide the data used for CARS into four types: users, items, contexts, and interactions. Following \cite{18}, we define context as the information that is associated with an interaction, \eg the current location, time, previous click, etc. Figure \ref{fig:dataProcessing} illustrates the data in CARS, where the main data is the user-item-context interaction tensor. In the sparse tensor, each nonzero entry $(u, i, c)$ denotes that the user $u$ has interacted with the item $i$ under the context $c$; we give such entries a label of 1, \ie $y_{uic}=1$. 
Each $u, i, c$ is respectively associated with a multi-hot feature vector $\textbf{u}$, $\textbf{i}$, and $\textbf{c}$, which contain the features that describe the user, item, and context. For example, $\textbf{u}$ includes static user profiles like gender and interested tags, $\textbf{i}$ includes static item attributes like category and price, and $\textbf{c}$ includes dynamic contexts like the current location of the user and the time. 

Given such data, we convert it to the form of attributed user-item bipartite graph that has the same representation power. Specifically, each vertex represents a user or an item, and each edge represents the interaction between the connected user and item. Each vertex or edge is associated with a feature vector $\textbf{u}$, $\textbf{i}$, or $\textbf{c}$. Note that there may exist multiple edges between a user-item pair, since a user may interact with the same item multiple times under different contexts. We denote all edges in the graph as the set $\mathcal{Y} = \{(u,i,c) | y_{uic}=1\}$, the neighbors of the user $u$ as the set $\mathcal{N}_u = \{ (i,c) | y_{uic}=1 \}$, and neighbors of the item $i$ as the set $\mathcal{N}_i = \{ (u,c) | y_{uic}=1 \}$. 

We formulate the problem of CARS as:
\begin{description}
\item[\textbf{Input}:] User-item-context interactions $\{ (u, i, c) | y_{uic}=1 \}$,  feature vectors of users  $\{ \textbf{u}\}$, items  $\{\textbf{i}\}$, and contexts  $\{\textbf{c}\}$.
\item[\textbf{Output}:] Prediction function $f: \textbf{u}, \textbf{i}, \textbf{c} \rightarrow \mathbb{R}$, which takes the feature vector of a user, an item, and a context as the input, and outputs a real value that estimates how likely the user will interact with the item under the context. 
\end{description} 

%% file: 4_gcm.tex
\section{Graph Convolution Machine (GCM)}

We present our method in this section. We first describe the predictive model, followed by the model complexity analyses and optimization details. 

\subsection{Predictive Model}
\begin{figure*}
\centering
\includegraphics[width=0.85\textwidth]{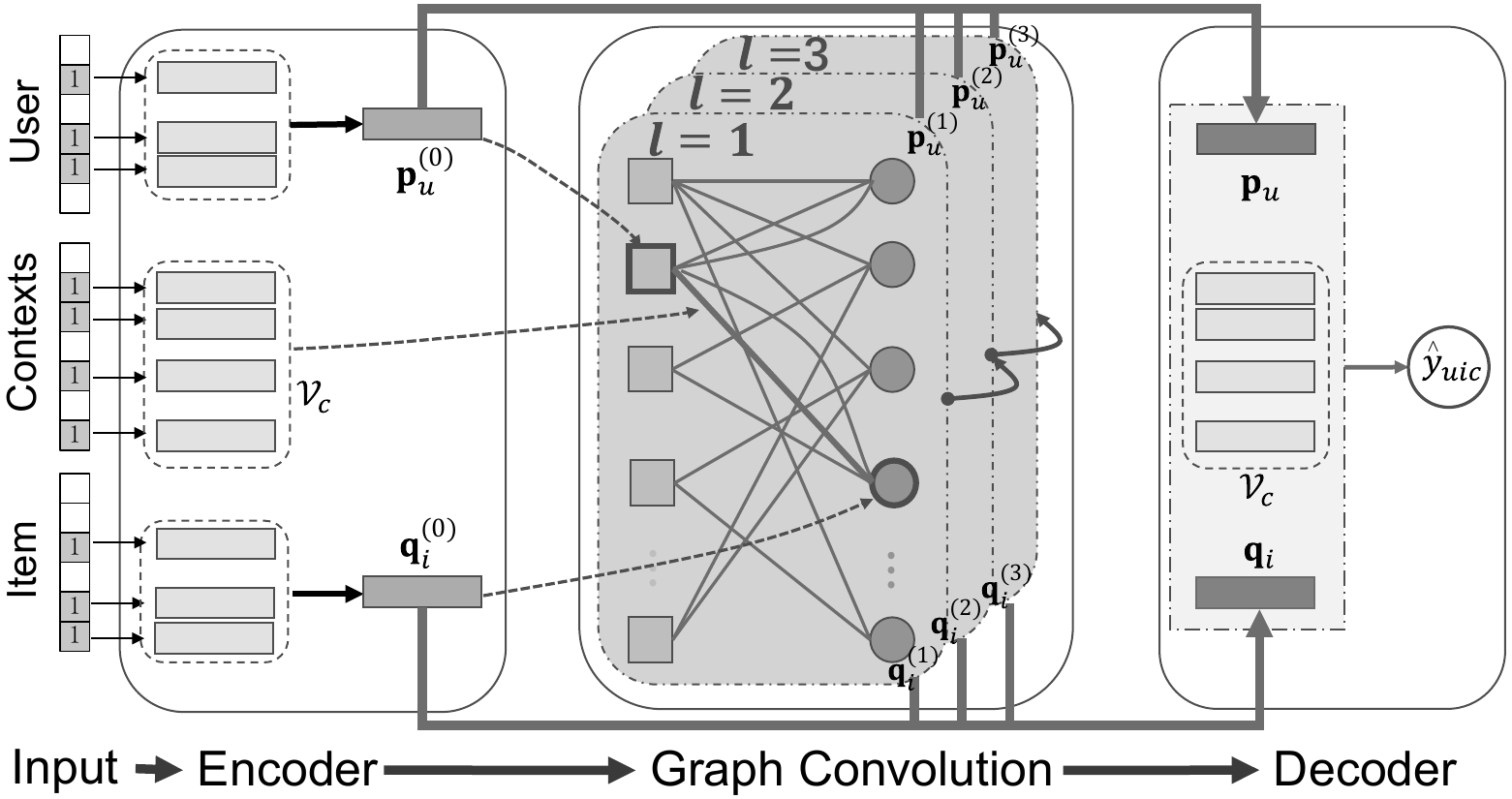}
\caption{The Graph Convolution Machine model.}
\label{fig:framework}
\end{figure*}

Figure \ref{fig:framework} illustrates the model framework, which consists of three components: an encoder, graph convolution layers, and a decoder. We next describe each component one by one. 

\subsubsection{Encoder}
The input to the encoder has three fields: user-field features $\textbf{u}$, item-field features $\textbf{i}$, and the context-field features $\textbf{c}$. 
We include the ID feature into the user-field and item-field features, since it helps to differentiate users (items) when their profiles (attributes) are the same\footnote{Note that there is no need to include ID into the context-field features, since a context $c$ and its features $\textbf{c}$ are one-to-one mapping.}. 
For each nonzero feature, we associate it with an embedding vector, resulting in a set of embeddings to describe the input user, item, and context, respectively. 
We then pool the set of user (and item) field into a vector, so as to feed the vector into the the following GC layers to refine the user (and item) representations. 
Specifically, we adopt average pooling, that is,

\begin{equation}
    \textbf{p}_u^{(0)} = \frac{1}{|\textbf{u}|} \textbf{P}^T \textbf{u},
\end{equation}
where $|\textbf{u}|$ denotes the number of nonzero features in $\textbf{u}$, and $\textbf{P}\in \mathbb{R}^{U\times D}$ is the embedding matrix for user features, where $U$ denotes the number of total user features and $D$ denotes the embedding size. 
$\textbf{p}_u^{(0)}$ denotes the initial representation vector for $u$. Similarly, we get the initial representation vector for item $i$ as $\textbf{q}_i^{(0)}$. 

Note that other pooling mechanisms can be applied here, such as the attention-based pooling~\cite{17,39,40} which learns varying weights for feature embeddings. However, we tried that and find it does not improve the performance. Thus we keep the simplest average pooling and avoid introducing additional parameters. 
Since we do not update the context representation in the following GC layers, we do not perform pooling on the context field. We denote the set of context-field embeddings as $\mathcal{V}_c = \{ \textbf{v}_s | s\in \textbf{c} \} $, where $s\in \textbf{c}$ denotes the nonzero feature in $\textbf{c}$ and $\textbf{v}_s$ denotes the embedding vector for context feature $s$. 
The encoder outputs $\textbf{p}_u^{(0)}$, $\textbf{q}_i^{(0)}$, and $\mathcal{V}_c$, which are fed into the next component of GC layers.  

\subsubsection{Graph Convolution Layers}
This is the core component of GCM, designed to address the limitation of existing supervised learning-based CARS models. 
It refines $\textbf{p}_u^{(0)}$ and $\textbf{q}_i^{(0)}$ by exploiting holistic user-item interaction data, which can augment the user and item representations with explicit collaborative filtering signal~\cite{3}. The GC on user-item graph is typically formulated as a message propagation framework:

\begin{equation}
    \textbf{p}_u^{(l+1)} = \sum_{i \in \mathcal{N}_u} g(\textbf{p}_u^{(l)},  \textbf{q}_i^{(l)}); \  \textbf{q}_i^{(l+1)} = \sum_{u \in \mathcal{N}_i} g(\textbf{q}_i^{(l)},  \textbf{p}_u^{(l)}), 
\end{equation}
where $\textbf{p}_u^{(l)}$ and $\textbf{q}_i^{(l)}$ denote the refined user representation and item representation of the $l$-th GC layer, respectively, and $g(\cdot)$ is a self-defined function. Recursively conducting such message propagation relates the representation of a user with her high-order neighbors, \eg first-order for interacted items and second-order for co-interacted users, which is beneficial for collaborative filtering; and the same logic applies to item representation. 

However, the standard GC does not consider the features on edges.
In our constructed user-item graph, the edges between a user and an item carry the context features, which are important to understand the context-dependent interaction patterns. 
For example, a user may prefer bars on Friday, and a restaurant is more popular on lunch time. As such, better user and item representations can be obtained if the context features can be properly integrated into the GC.  

To this end, we propose a new GC operation that incorporates the edge features of contexts:

\begin{equation}
\begin{split}
    \mathbf{p}_u^{(l+1)} = \sum_{(i,c)\in \mathcal{N}_u} \dfrac{1}{\sqrt{|\mathcal{N}_u|}}\big( \mathbf{q}_i^{(l)} + \frac{1}{|\mathcal{V}_c|} \sum_{\textbf{v}_s\in \mathcal{V}_c} \textbf{v}_s \big), \\
    \mathbf{q}_i^{(l+1)} = \sum_{(u,c)\in \mathcal{N}_i} \dfrac{1}{\sqrt{|\mathcal{N}_i|}}\big( \mathbf{p}_u^{(l)} + \frac{1}{|\mathcal{V}_c|} \sum_{\textbf{v}_s\in \mathcal{V}_c} \textbf{v}_s \big).
\end{split}
\label{eq:gc}
\end{equation}
Next we explain the rationality of the GC of the user side, since the item side can be interpreted in the same way. 
Here $|\mathcal{N}_u|$ denotes the number of edges connected with the user $u$, and the coefficient $\frac{1}{\sqrt{|\mathcal{N}_u|}}$ is a normalization term to avoid the scale of embedding values increasing with the GC.
We incorporate the context features by averaging their embeddings and adding to the connected user embedding. 
Through this way, we build the connection between a user with both her interacted item and the interacted context. 
It is expected to capture the effect that if a user likes to choose an item under a certain context, then the similarity among their representations is similar. 
Note that we have tried more complicated mechanisms like incorporating the pairwise interactions among $\mathcal{V}_c$ and $\textbf{q}_i^{(l)}$, and using a MLP to capture high-order interactions. 
However these ways do not lead to performance improvements. Thus we use this simple average operation, which is easy to interpret and train (no additional parameters are introduced).  

By stacking multiple such GC layers, a user (or an item) representation can be refined by its multi-hop neighbors. Since the representation of different layers carry different semantics, we next combine the representations of all layers to form a more comprehensive representation:

\begin{equation}
    \mathbf{p}_u = \sum_{l=0}^L \alpha_l \mathbf{p}_u^{(l)}; \quad  \mathbf{q}_i = \sum_{l=0}^L \alpha_l \mathbf{q}_i^{(l)},
\end{equation}
where $\alpha_l$ denotes the weight of the $l$-th layer representation, 
which can be treated as hyper-parameter and tuned via a grid search with the constraint that $\alpha_l \geq 0$ and $\sum_{l=0}^L \alpha_l = 1$.
However, the workload of tuning them increases exponentially, as the GCN goes deep.
In our experiments, we find that setting $\alpha_l$ to $1/(L+1)$ leads to satisfactory performance in general. Therefore, we fix $\alpha_l$ to $1/(L+1)$ for simplicity.
A possible extension is to learn $\alpha_l$, \eg designing attention mechanism or optimizing them on the validation data. We leave this extension as future work, since it is not the focus of this work. 

In what follows, we provide the matrix form of GC layers for implementation. Let the user-item interaction matrix be $\mathbf{R}_{\mathrm{u}\mathrm{i}} \in \mathbb{R}^{N \times M}$, where $N$ and $M$ denotes the number of users and items. Each entry $r_{ui} \in \mathbf{R}_{\mathrm{u}\mathrm{i}}$ is the number of times user $u$ interacts with item $i$. 
Similarly, we utilize  $\mathbf{R}_{\mathrm{u}\mathrm{c}} \in \mathbb{R}^{N \times K}$ and $\mathbf{R}_{\mathrm{i}\mathrm{c}} \in \mathbb{R}^{M \times K}$  to denote user-context interaction matrix and item-context interaction matrix respectively, where $K$ is the number of contexts. Then we define the adjacency matrix of user-item-context graph as

\begin{equation}
    \mathbf{A} = 
    \begin{pmatrix}
        \mathbf{0} & \mathbf{R}_{\mathrm{u}\mathrm{i}} & \mathbf{R}_{\mathrm{u}\mathrm{c}} \\
        \mathbf{R}_{\mathrm{u}\mathrm{i}}^T & \mathbf{0} & \mathbf{R}_{\mathrm{i}\mathrm{c}} \\
        \mathbf{0} & \mathbf{0} & 2\mathbf{I}
    \end{pmatrix},
\end{equation}
where $\mathbf{0}$ is all-zero matrix, $\mathbf{I}$ is identity matrix. Let $\mathbf{D}$ be diagonal degree matrix of $\mathbf{A}$, that is, the $t$-th diagonal element $\mathbf{D}_{tt}=\sum_j \mathbf{A}_{tj}$. The normalized adjacency matrix can be expressed as

\begin{equation}
    \hat{\mathbf{A}} = \sqrt{2}\mathbf{D}^{-\frac{1}{2}}\mathbf{A}.
\end{equation}
Then, we get the matrix form of the layer-wise propagation rule which is equivalent to Equation~\eqref{eq:gc}:

\begin{equation}
    \mathbf{E}^{(l)} = \hat{\mathbf{A}} \mathbf{E}^{(l-1)},
\end{equation}
where $\mathbf{E}^{(l)} \in \mathbb{R}^{(N+M+K)\times D}$ is the concatenate of user, item and context embedding matrix. $\mathbf{E}^{(0)}$ is set as the concatenate matrix of encoded embedding tables from Encoder, which can be expressed as

\begin{equation}
    \mathbf{E}^{(0)} = [ \underbrace{\mathbf{p}_{u_1}^{(0)}, \cdots \mathbf{p}_{u_N}^{(0)} }_{\text{user embeddings}}, \underbrace{\mathbf{q}_{i_1}^{(0)}, \cdots \mathbf{q}_{i_M}^{(0)} }_{\text{item embeddings}}, \underbrace{\mathbf{r}_{c_1}, \cdots \mathbf{r}_{c_K} }_{\text{context embeddings}} ]^T.
\end{equation}
Lastly, we get the final embedding matrix

\begin{align}
\begin{split}
\mathbf{E} 
    &= \alpha_0 \mathbf{E}^{(0)} + \alpha_1 \mathbf{E}^{(1)} + \alpha_2 \mathbf{E}^{(2)} \cdots + \alpha_L \mathbf{E}^{(L)} \\
    &= \alpha_0 \mathbf{E}^{(0)} + \alpha_1\hat{\mathbf{A}} \mathbf{E}^{(0)} + \alpha_2\hat{\mathbf{A}}^2 \mathbf{E}^{(0)} + \cdots + \alpha_L\hat{\mathbf{A}}^L \mathbf{E}^{(0)}.
\end{split}
\end{align}

\subsubsection{Decoder}
The GC layers output refined representation of user $\textbf{p}_u$ and item $\textbf{q}_i$, and keep the embeddings of context features unchanged. 
The role of the decoder is to output the prediction score by taking in the representations. The standard choice of decoder is multi-layer perceptron (MLP),
which however falls short here since it only models feature interactions in an implicit way. 
In CARS, explicitly modeling the interactions between features is known to be important for user preference estimation~\cite{14}. 
For example, the classic factorization machine (FM) models the pairwise interactions between feature embeddings and has long been a competitive model for CARS. 

Inspired by the simplicity (linear model) and the effectiveness of FM, we adopt it as the decoder of GCM. 
The idea is to explicitly model the pairwise interactions between the (refined) representations of user, item, and contexts with inner product. 
Specifically, let the set of vectors $\mathcal{V}$ be $\mathcal{V}_c \cup \{\textbf{p}_u, \textbf{q}_i$\}, the decoder outputs the prediction score as:

\begin{equation}
    \hat{y}_{uic} = \frac{1}{2} \left(\sum_{\textbf{v}_s \in \mathcal{V}} \sum_{\textbf{v}_t \in \mathcal{V}} \textbf{v}_s^T \textbf{v}_t - \sum_{\textbf{v}_s \in \mathcal{V}} \textbf{v}_s^T \textbf{v}_s \right).
\end{equation}
Here the self-interactions $\textbf{v}_s^T \textbf{v}_s$ are excluded since they are useless for the prediction. The bias terms for each user, item, and context feature are omitted for clarity. 

Note that our FM-based decoder slightly differs from the vanilla FM, which models the interactions between the embeddings of all input features. Here we project each user (item) into a vector, rather than retaining the embeddings of her (its) features. 
An advantage is that this way abandons the internal interactions of user-field (item-field) features, shedding more light on the interactions between user (item) and context features, which is as expected. 

\subsection{Model Complexity Analyses}
We analyze the complexity of GCM from two aspects: the number of trainable parameters and the time complexity. 

All trainable parameters come from the encoder layer, \ie the embeddings of input features, since the GC layers and the decoder layer introduce no parameters to train. Let the feature number for the user field, item field, and context field as $U, I,$ and $C$, respectively, and the embedding size be $D$. Then the embedding layer costs $(U+I+C)\times D$ parameters. This demonstrates the low model complexity of GCM, since the number of trainable parameters is the same as FM --- the most simple embedding-based CARS model. 

For model training, since the complexity of the encoder plus the decoder is the same as that of FM, we analyze the additional time complexity caused by the GC layers. 
We implement the training in the batch-wise matrix form. 
Assume a batch contains all interactions. 
Then performing one GC layer takes time $O( (|\mathcal{Y} |+N+M )D )$, where $N$ and $M$ denote the number of users and items, respectively. This complexity increases linearly with the number of GC layers. 

After the model is trained, we perform one pass of GC layers to obtain the refined representations of all users and items, which can be done offline before online serving. 
As such, during online serving, we only need to execute the decoder, which has the same time complexity of FM. 
This is much faster than the recently emerging deep neural network-based CARS models like xDeepFM~\cite{16} and Convolutional FM~\cite{17}.
Table \ref{tab:serving-time} shows the model inference time of evaluating 1000 Yelp-OH users in which each interaction has 10 nonzero features of embedding size 64 and batch size is 4000. The testing platform is GeFore GTX 1080Ti with 16GB memory CPU. As can be seen, GCM takes similar time as FM, being 24.5 and 157.7 times faster than xDeepFM and Convolutional FM, respectively.

\begin{table}[]
\centering
\caption{Model inference time of evaluating 1,000 Yelp-OH users (14 million interactions and 10 nonzero features per interaction).}
\resizebox{0.45\textwidth}{!}{
\begin{tabular}{@{}cccccc@{}}
\toprule
Model & FM & GCM & GIN & xDeepFM & Convolutional FM \\ \midrule
Time/s & 8.51 & 14.93 & 35.45 & 365.82 & 2354.25 \\ \bottomrule
\end{tabular}}
\label{tab:serving-time}
\end{table}

\subsection{Optimization}
To optimize model parameters, we opt for the pointwise log loss, which is a common choice in recommender system~\cite{1,16}. In each training epoch, we randomly sample non-observed interactions for each instance in $\mathcal{Y}$ to form the negative set $\mathcal{Y}^{-}$ --- that is, for each observed instance $(u,i,c) \in \mathcal{Y}$ of Yelp (or Amazon) dataset, we randomly match 4 (or 2) items from the item pool that user $u$ has not interacted under context $c$. Then we minimize the following objective function:
\begin{equation}\small
    L= -\sum_{(u,i,c)\in\Set{Y}} \log{\sigma(\hat{y}_{uic})} - \sum_{(u,i,c)\in\Set{Y}^-}\log{(1-\sigma(\hat{y}_{uic}))}+\lambda\norm{\Theta}^{2}_{2},
\end{equation}
where $\sigma(\cdot)$ is the sigmoid function, $\lambda$ controls the $L_2$ regularization to prevent over-fitting. The optimization is done by mini-batch Adam~\cite{41}. 

%% file: 5_experiment.tex
\section{Experiments}
We evaluate experiments on three benchmark datasets, aiming to answer the following research questions:
\begin{itemize}
    \item \textbf{RQ1:} Compared with the state-of-the-art models, how does GCM perform \wrt top-$k$ recommendation?
    \item \textbf{RQ2:} How do different settings (\eg depth of layer, modeling of context features, design of decoder) affect GCM?
    \item \textbf{RQ3:} How do the representation learning benefit from multiple interactions among users, items and contexts for item cold start issue?
\end{itemize}

\subsection{Experimental Settings}
\subsubsection{Dataset Description}
To demonstrate the effectiveness of GCM, we conduct experiments on three datasets from Yelp and Amazon, which are publicly available and vary in domain and size. We summarize the statistics of datasets in Table~\ref{tab:data_stats}.
\begin{itemize}
    \item \textbf{Yelp}: This dataset is released by Yelp and records users' reviews on local businesses like bars and restaurants. In particular, we extract records happened in two different areas of USA --- North Carolina, Ohio States --- to construct datasets, termed Yelp-NC and Yelp-OH respectively.
    \item \textbf{Amazon}: Amazon review data is widely used in recommendation \cite{3}. We select book subset from the collection in this work, and term it Amaon-book.
\end{itemize}

\begin{table}[t]
\caption{Statistics of the datasets. We omit ID feature when counting the number of user and item features}
\centering
\label{tab:data_stats}
\resizebox{0.45\textwidth}{!}{
\begin{tabular}{lrrr}
\toprule
Dataset           & \multicolumn{1}{l}{Yelp-NC} & \multicolumn{1}{l}{Yelp-OH} & \multicolumn{1}{l}{Amazon-book} \\
\midrule
\#User            & 6,336                      & 5,170                      & 44,709                     \\
\#Item            & 13,003                      & 12,997                      & 46,831                     \\
\#Instance        & 185,408                     & 143,884                     & 1,174,785                   \\
\#User Feature    & 24                          & 24                          & -                          \\
\#Item Feature    & 68                          & 213                         & 24,816                          \\
\#Context Feature & 13,209                      & 13,347                      & 46,900      \\                  
\bottomrule
\end{tabular}}
\end{table}

In what follows, we briefly introduce the features of users, items, and contexts.
Specifically, for Yelp-NC and Yelp-OH, each user profile includes \textit{yelping\_since\_year}\footnote{We only keep the \textit{year} of the \textit{yelping\_since} field which indicates the time the user joined Yelp.} and \textit{average\_stars}, while the pre-existing features of items are composed of three attributes: \textit{city}, \textit{stars} and \textit{is\_open}.
We treat each review record as an observed instance, and collect \textit{city}\footnote{
The context feature \textit{city} means which city does the interaction happen on. It is set as the city of the interacted item.
}, \textit{month}, \textit{hour}, \textit{day\_of\_week} and \textit{last\_purchase} as its context feature.
For Amaon-book, the static features of items are composed of two attributes: \textit{price} and \textit{brand}. Similarly, each review record is treated as an observed instance, and \textit{year}, \textit{month}, \textit{day}, \textit{day\_of\_week} and \textit{last\_purchase} are collected as its context feature.
Moreover, for all datasets, the 10-core setting is adopted to ensure data quality, \ie retaining users with at least ten interactions.

For each user, we select the last interaction record to constitute the test set, while the remains are served as the training set.
To emphasize model capability in recommending novel items for a user, we further filter the training set if the user-item pairs have appeared in the test set.

\subsubsection{Evaluation Metrics}
In the evaluation phase, for each user in the test set, we view all items that she has not consumed before as recommendation candidates.
Each method outputs a ranking list over the candidates.
We then adopt two widely-used protocols to evaluate the quality of ranking lists: Hit Ratio (HR) and Normalized Discounted Cumulative Gain (NDCG).
In particular, HR@$K$ measures whether the test item is in the top-$K$ positions of the recommended list, whereas NDCG@$K$ assigns higher scores to the top-ranked items. In our experiments, we report the results of $K=10$ and $K=50$.

\subsubsection{Baselines}

We compare our GCM with several methods as follows:
\begin{itemize}
    \item \textbf{MF}~\cite{42}: This exploits the user-item interactions only to learn user and item embeddings, while forgoing the context features.
    
    \item \textbf{LightGCN}~\cite{43}: Such model is the state-of-the-art GNN-based CF recommender, which incorporates high-order connectivity in user-item interaction graph into embeddings, while neglecting context features.

    \item \textbf{FM}~\cite{11}: This takes into account all information related to an interaction by converting all information to a feature vector then modeling second-order feature interaction to predict user preference.
    
    \item \textbf{NFM}~\cite{14}: This model leverages a MLP to capture nonlinear and high-order interaction among user, item, and context features.
    
    \item \textbf{xDeepFM}~\cite{16}: This is a recent neural FM model which combines explicit and implicit high-order feature interactions.

    \item \textbf{GIN}~\cite{32}: This is a graph-based model which mines user intention by applying implicit intention propagation and attention mechanism on commodity similarity graph.
\end{itemize}

\begin{table*}[t]
\caption{Overall Performance Comparison. The bold indicates the best result, while the second-best performance is underlined.}
\label{tab:overall-performance}
\resizebox{0.99\textwidth}{!}{
\begin{tabular}{lcccc|cccc|cccc}
\toprule
\multirow{3}{*}{} & \multicolumn{4}{c|}{\textbf{Yelp-NC}}                                 & \multicolumn{4}{c|}{\textbf{Yelp-OH}}                                 & \multicolumn{4}{c}{\textbf{Amazon-book}}                                       \\
                                & \multicolumn{2}{c}{HR}            & \multicolumn{2}{c|}{NDCG}         & \multicolumn{2}{c}{HR}            & \multicolumn{2}{c|}{NDCG}         & \multicolumn{2}{c}{HR}            & \multicolumn{2}{c}{NDCG}          \\
                                & @10             & @50             & @10             & @50             & @10             & @50             & @10             & @50             & @10             & @50             & @10             & @50             \\ \hline\hline
MF                              & 0.0384          & 0.1173          & 0.0175          & 0.0341          & 0.0429          & 0.1261          & 0.0206          & 0.0383          & 0.0402          & 0.1243          & 0.0203          & 0.0382          \\
LightGCN                        & 0.0499          & 0.1394          & 0.0241          & 0.0431          & 0.0518          & 0.1520          & 0.0249         & 0.0461          & 0.0543          & 0.1466          & 0.0274          & 0.0473          \\
FM                              & 0.0739          & 0.1804          & 0.0396          & 0.0624          & 0.1959          & 0.4201          & 0.1049          & 0.1538          & 0.0587          & 0.1477          & 0.0323          & 0.0514          \\
NFM                             & 0.0824          & 0.2110          & 0.0419          & 0.0695          & 0.2248          & 0.4836          & 0.1161          & 0.1725          & 0.0808          & 0.1954          & 0.0444          & 0.0692          \\
xDeepFM                         & 0.0851          & 0.2086          & {\ul 0.0458}    & {\ul 0.0723}    & 0.2296          & 0.4799          & 0.1218          & 0.1762          & 0.0886          & 0.2119          & 0.0481          & 0.0748          \\
GIN                             & {\ul 0.0866}    & {\ul 0.2175}    & 0.0449          & 0.0722          & {\ul 0.2304}    & {\ul 0.4965}    & {\ul 0.1238}    & {\ul 0.1818}    & {\ul 0.0939}    & {\ul 0.2189}    & {\ul 0.0502}    & {\ul 0.0774}    \\
GCM                             & \textbf{0.1046} & \textbf{0.2421} & \textbf{0.0557} & \textbf{0.0854} & \textbf{0.2648} & \textbf{0.5166} & \textbf{0.1457} & \textbf{0.2008} & \textbf{0.0968} & \textbf{0.2232} & \textbf{0.0536} & \textbf{0.0810} \\ \hline\hline
\%Improv.                       & 20.78\%         & 11.31\%         & 21.62\%         & 18.12\%         & 14.93\%         & 4.05\%          & 17.69\%         & 10.45\%         & 3.08\%          & 1.96\%          & 6.77\%          & 4.65\%          \\
$p$-value                         & 3.35e-9         & 4.37e-7         & 6.75e-10        & 5.91e-10        & 5.36e-12        & 9.10e-6         & 8.22e-9         & 2.45e-8         & 1.86e-4         & 3.41e-4         & 1.70e-3         & 1.05e-3         \\
\bottomrule
\end{tabular}}
\end{table*}

Fi-GNN~\cite{31} is a recent work on click-through rate prediction with graph neural network, which is highly relevant with our work. It differs from GCM in graph construction --- it builds a feature graph for each interaction, rather than the user-item graph. As a graph needs be built for each interaction to obtain its prediction, the method is very slow in evaluation since all recommendation candidates need be scored. As such, this method is not suitable for our all-ranking CARS evaluation, and we do not further compare with it. The Convolutional FM is not compared for the same reason (see Table \ref{tab:serving-time} for model inference time). 

\subsubsection{Parameter Settings}
We implement our GCM model and all baselines in Tensorflow and will release our code upon acceptance.
We apply the mini-batch Adam to optimize all models, the learning rate and batch size are set to $0.001$ and 2048 respectively.
A grid search is conducted for confirming optimal hyperparameters: 
for LightGCN, the number of gcn layers is searched in $\{1,2,3,4\}$; 
for NFM, the number of hidden layers is set to 1, the dropout rate is tuned in $\{0.9,0.8,\cdots,0.1\}$ for bi-interaction layer and hidden layer respectively; 
for xDeepFM, the number of cross layers is searched in $\{1,2,3\}$ with neuron number per layer in $\{10,20,50,100,200\}$, the number of DNN layers is same to that of cross layers, while the neuron number per layer is set to 100; 
for GIN, the length of previous records is 1 since we only keep the last purchase information in the datasets, the depth parameter is searched in $\{1,2,3,4\}$, the number of neighbor nodes is tuned in $\{10,20\}$, the neighbor is selected by the Top-N function according to the edge weight (for nodes with few neighbors, we randomly sample from unconnected nodes as their potential neighbors), a 5-layer full-connection perception with ReLU activation is adopted as the setting in \cite{32}; 
for the proposed GCM, we search the model depth $L$ amongst $\{1,2,3\}$ with $\alpha_l=1/(L+1)$, and adopt average pooling to generate the final refined representations of GC layers. 
For all models, the coefficient of $L_{2}$ regularization term is searched in $\{10^{-5},10^{-4},10^{-3},10^{-2},10^{-1}\}$.
Moreover, we set the embedding size to 64 and train all models from the scratch.

\subsection{Performance Comparison (RQ1)}

We report the empirical results of all models in Table~\ref{tab:overall-performance} and have the following observations:
\begin{itemize}
    \item Clearly, MF achieves the worst performance on three datasets, indicating that modeling user-item pairs as isolated instances limits the representation ability severely. LightGCN obtains consistent improvements over MF. We attribute such improvements to the modeling of user-item connectivity. However, neither MF nor LightGCN takes the context features into consideration, ignoring important factors and being insufficient for CARS.

    \item FM, NFM and xDeepFM consistently outperform MF and LightGCN across all cases. This is reasonable since they incorporate context features into the representation learning, so as to achieve better expressiveness and help to solve the data sparsity issue; Among them, NFM and xDeepFM perform better than FM by a large margin since they model more complex feature interactions: NFM employs MLP on user, item, and context features to capture their nonlinear and complex interactions, while xDeepFM learns high-order feature interactions in a more explicit way through a CIN network. This verifies that simply linear functions (\eg inner product adopted by MF and LightGCN) might limit the representation learning and interaction modeling.
    
    \item GIN is the strongest baseline in all cases except for NDCG@10 and NDCG@50 in Yelp-NC. Such improvements is mainly because of GIN's capability to model user intention by applying message propagation in commodity similarity graph, which also verify the necessity of bridging the relationship among data instances.    
    
    \item GCM consistently outperforms all baselines \wrt all measures. In particular, GCM achieves noticeable improvements over the strongest baselines \wrt HR@10 by 20.78\%,  14.93\%, and 3.08\%, in Yelp-NC, Yelp-OH, and Amazon-book, respectively.
    From t-test, we can find $p$-value $< 0.05$ across the board, indicating that the improvements of GCM over the strongest baseline are statistically significant.
    We attribute such improvements to:
    1) GCM employs the embedding propagation over the attributed graph to distill useful information from neighbors and connected edges, thus improving the representation ability; 2) Comparing with GIN which only propagates item embedding in the graph, GCM integrates the representations of users, items and contexts into the graph for information propagation, which may results in a more unified representations; and 3) Having established the refined representations, GCM further adopts FM to explicitly model the feature interactions.
\end{itemize}

\subsection{Study of GCM (RQ2)}
We next report ablation studies to verify the rationality of some designs in GCM, 
\ie analyzing the influence of model depth, context modeling, normalization term, and decoder.

\subsubsection{Impact of Model Depth}
\begin{figure}
    \subfigure[Yelp-NC]{\includegraphics[width=0.46\linewidth]{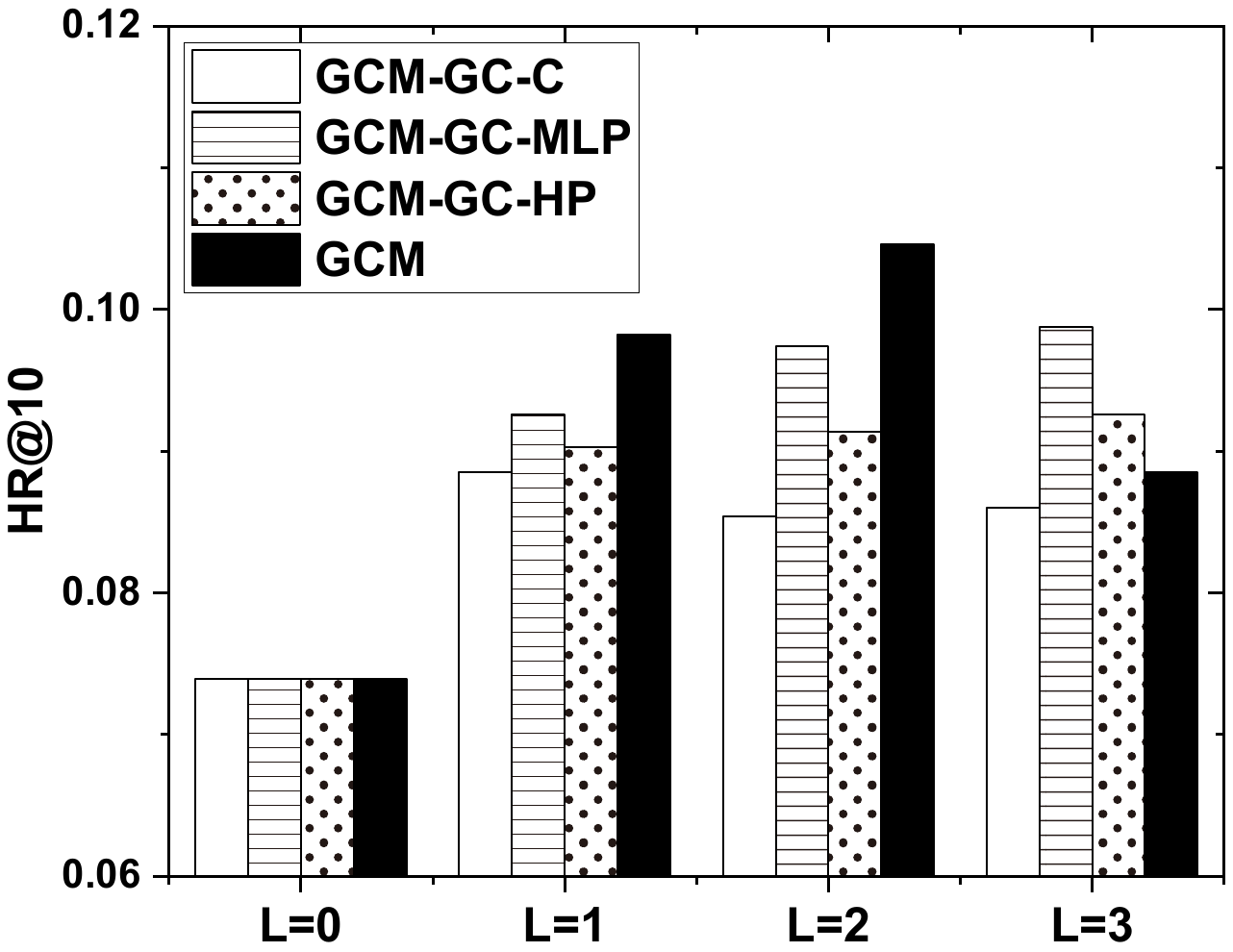}}
    \subfigure[Yelp-OH]{\includegraphics[width=0.46\linewidth]{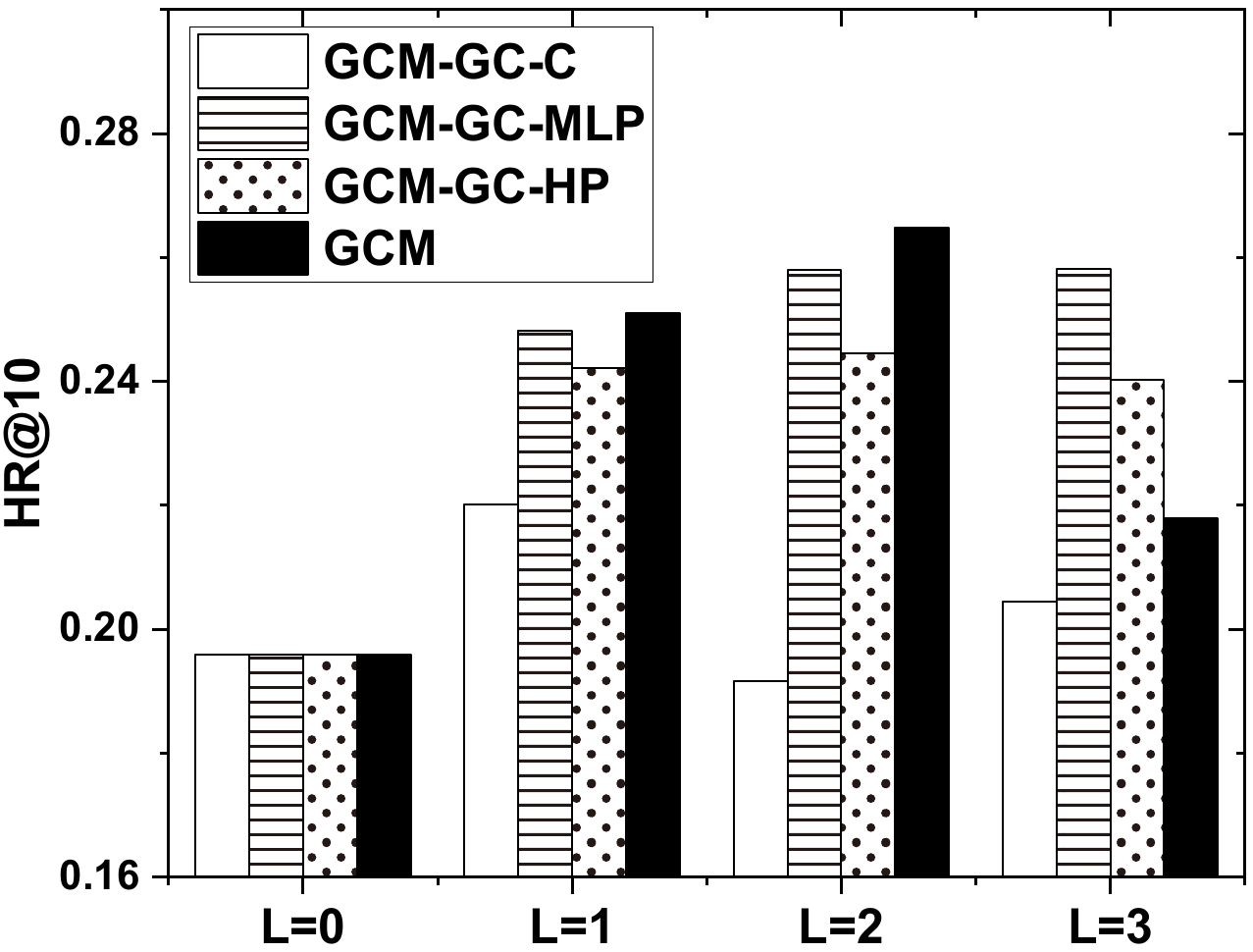}}
    \caption{The impact of depth and propagation rule in GC.}
    \label{fig:ablation}
\end{figure}

As GC is the core of GCM and stacking more GC layers is expected to augment the user and item representations with information propagated from multi-hop neighbors, we investigate how the number of GC layers affects the performance.
In particular, we search the number of GC layers, $L$, in the range of $\{0,1,2,3\}$ and report the empirical results in Figure~\ref{fig:ablation}.

We use GCM-1 to represent the model with one GC layer, and similar notations for others.
We have several findings:
\begin{itemize}
    \item GCM-0 disables the embedding propagation over user-item attributed graph and downgrades to a FM-like linear model, thereby achieving poor performance. This again justifies the importance of GC layers.
    
    \item Obviously, increasing the number of GC layers results in better performance from $L=0$ to 2. In particular, GCM-2 performs better than GCM-1 in both datasets. It is reasonable since the signals passing from multi-hop neighbors (\eg the second-order connectivity between behaviorally similar users or co-purchased items) are encoded into user and item representations of GCM-2, while GCM-1 only exploits personal history to enrich representations. This observation is consistent with that in NGCF~\cite{3}.
    We also tried to stack more GC layers (\ie GCM-3), finding improvement degrades and over-smoothing issue. This suggests that GCM benefits from the first- and second-order neighbors most, but may suffer from degradation when higher-order neighbors are involved.
    
\end{itemize}

\subsubsection{Impact of Context Modeling}
One major contribution of GCM is to organize the context features as edges in the attributed user-item graph.
We hence perform ablation study, to demonstrate the rationality and effectiveness of this design.
In particular, we build three different propagation rules for the GC layers of GCM: 1) GCM-GC-C removes the context features from the attributed graph and keeps the vanilla user-item interaction graph only; 2) GCM-GC-MLP first replaces the addition operation with concatenation in Equation~\eqref{eq:gc}, then generates the message vector through a MLP; 3) GCM-GC-HP encodes the Hadamard product of the representations of neighboring node and their connected edge into the message during message passing.
We show the comparison results in Figure~\ref{fig:ablation} and have the following observations.
\begin{itemize}
    \item Modeling context features as the edges endows GCM with better generalization ability.
    In particular, GCM-GC-C performs worst among four competing methods in all cases, demonstrating the necessity of modeling context features when performing message propagation.
    
    \item GCM-GC-MLP consistently achieves better recommendation accuracy than GCM-GC-HP. One possible reason is that equipped with MLP, GCM-GC-MLP can model non-linear and high-order feature interactions, resulting in better representation ability.
    
    \item On both datasets, the best performance is always achieved by 2-layers GCM, which again justifies the rationality of our context modeling strategy. Mathematically, GCM can be viewed as a special case of GCM-GC-MLP in which the learnt weights are the concatenation of two identity matrix. However, as more trainable parameters are involved, it would require more data to learn the function which may encounter difficulty in actually learning process, especially for the notorious problem of data sparsity in recommendation system.~\cite{44}.
    
    \item Jointly analyzing Table~\ref{tab:overall-performance} and Figure~\ref{fig:ablation}, we find that GCM-GC-C without considering contexts achieves better performance than other baselines in Yelp-NC and comparable performance in Yelp-OH. This empirically suggests that propagating embeddings over interaction graphs is of importance to generate high-quality representations.
    
\end{itemize}

\subsubsection{Impact of normalization term}
For convenience, we only present the variants of the GC of the user side, since the same logic can be applied to the item side.
In GCM, we employ sqrt normalization term $\frac{1}{\sqrt{|\mathcal{N}_u|}}$ on each neighbor embedding when performing neighborhood aggregation. To verify its rationality, we explore two different variants and report their empirical results here.
The first variant uses symmetric normalization term, \ie $\frac{1}{\sqrt{|\mathcal{N}_u|}\sqrt{|\mathcal{N}_i|}}$, which is a common choice in GCN-based models~\cite{43}, we term it GCM-sym. The other variant uses $L_1$ normalization, \ie $\frac{1}{|\mathcal{N}_u|}$, we term it GCM-$L_1$. Table~\ref{tab:norm and decoder impact} shows the results of the 2-layer GCM. We have the following observations:
\begin{itemize}
    \item The best setting in general is using sqrt normalization term on single side (\ie the current degign of GCM). Adding additional regularization coefficients will greatly drop the performance.
    \item The smaller the normalization term, the worse the performance. To understand this observation, we can see the following inequalities: $\frac{1}{\sqrt{|\mathcal{N}_u|}} > \frac{1}{\sqrt{|\mathcal{N}_u|}\sqrt{|\mathcal{N}_i|}} > \frac{1}{\sqrt{|\mathcal{N}_u|}\sqrt{|\mathcal{N}_u|}} = \frac{1}{|\mathcal{N}_u|}$. The second inequality is due to $|\mathcal{N}_u| > |\mathcal{N}_i|$ on everage in Yelp-NC and Yelp-OH.
\end{itemize}

\subsubsection{Impact of Decoder}
\begin{table}[]
\centering
\caption{The variants of GCM with different normalization terms and decoders}
\resizebox{0.45\textwidth}{!}{
\begin{tabular}{lcccc}
\toprule
\multirow{2}{*}{} & \multicolumn{2}{c}{\textbf{Yelp-NC}} & \multicolumn{2}{c}{\textbf{Yelp-OH}} \\
                                & HR@10             & NDCG@10           & HR@10            & NDCG@10           \\ \hline
GCM                             & 0.1046            & 0.0557            & 0.2648           & 0.1457            \\ \hline
GCM-$L_1$                       & 0.0810            & 0.0421            & 0.2373           & 0.1246            \\
GCM-sym                         & 0.0994            & 0.0527            & 0.2507           & 0.1383            \\ \hline
GCM-APC                          & 0.0947            & 0.0497            & 0.2321           & 0.1265            \\
GCM-MLP                         & 0.0892            & 0.0458            & 0.2263           & 0.1211            \\
GCM-MF                          & 0.0497            & 0.0253            & 0.0520           & 0.0251            \\ \bottomrule
\end{tabular}}
\label{tab:norm and decoder impact}
\end{table}
Having applied GC layers, we equip GCM with a decoder to model the pairwise interactions between the refined representations of users, items, and contexts.
Here we investigate the role of such decoder.
Towards this end, we compare GCM with three variants: 1) GCM-APC, that performs average pooling on context features before pair-wise inner product; 2) GCM-MLP, that replaces the decoder with MLP; and 3) GCM-MF, that replaces FM with inner product on user and item representations.
Table~\ref{tab:norm and decoder impact} shows the comparison of results. There are several observations:
\begin{itemize}
    \item Clearly, modeling feature interactions in the decoder enhances the predictive results. In particular, GCM, GCM-APC and GCM-MLP perform consistently better than GCM-MF, which relies only on the inner product of user and item representations. 
    
    \item While having encoded context features into user and item representations via GC layer, GCM and GCM-APC highlight their influence in an explicit fashion, while GCM-MLP models the feature interactions in a rather implicit way. The better performances of GCM and GCM-APC again verify the rationality and effectiveness of FM-based decoder.
    In addition, after performing average pooling on context features, GCM-APC degenerates the performance by a noticeable margin, the reason is that GCM endows higher weights on feature interaction between context field and user (item) field, which is the core of CARS.
    
\end{itemize}

\subsection{Performance \wrt Item Popularity (RQ3)}
\begin{figure}
    \subfigure[Yelp-NC]{\includegraphics[width=0.92\linewidth]{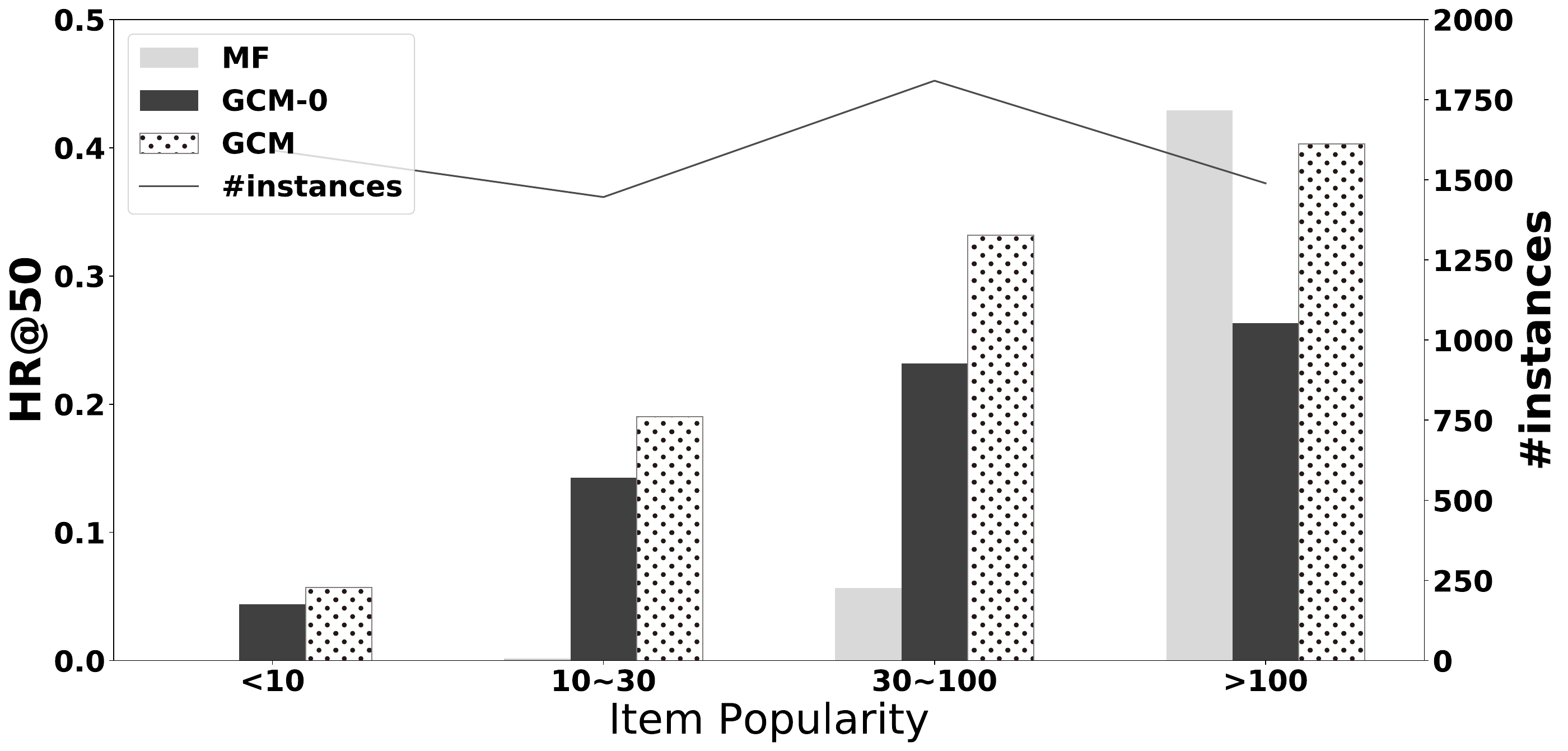}}
    \subfigure[Yelp-OH]{\includegraphics[width=0.92\linewidth]{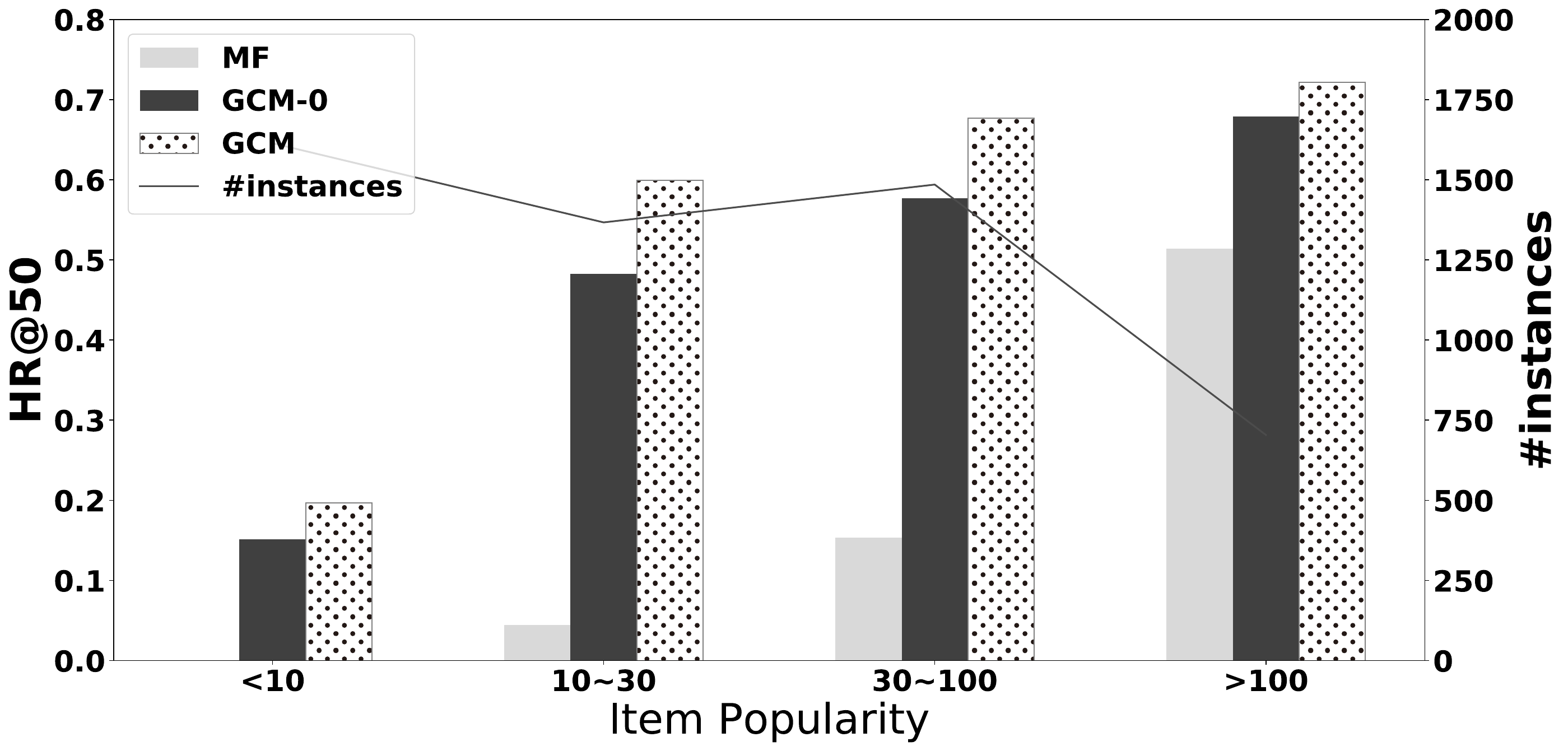}}
    \caption{Performances with respect to item popularity.}
    \label{fig:sparse}
\end{figure}

To alleviate the issue of item cold start of CF models, taking side information into account is an auxiliary strategy go beyond modeling user-item interaction. In the proposed GCM, We apply gc layers to capture high-order connectivity on user-item graph, which breaks down the independent interaction assumption of non-graph-based methods. We argue that such connectivity is a potential side information for cold-start issue. 
To verify this viewpoint, we split the test set according to the popularity (the number of interaction records) of the target item, and report the performance of MF~\cite{42}, GCM-0 and GCM in Figure~\ref{fig:sparse}. We have the following observations:

\begin{itemize}
    \item MF performs poorly at unpopular items, which indicates the item cold-start issue for CF models. GCM-0 has significant improvements in recommending uncommon items by introducing side information and modeling feature interactions. Our GCM can further improve performance by 20\%-30\%. We attribute such improvements to modeling high-order connectivity since gnn increases the possibility of unpopular item being exposed through high-order links, thereby expanding the training data of unpopular items.
    \item For popular items, MF achieves comparable performance with GCM, even prevails over GCM in Yelp-NC. The possible reason is that the data of popular items occupies the majority of the training data, making MF adopt a cautious strategy --- biased to recommending generally accepted items. Instead, GCM will recommend items that are more niche but still consistent with user's taste.
    \item The gain brought by gcn decreases as the popularity of items increases. This shows that as the number of neighbors increases, gcn may suffer from over-smoothing, since these items have too many audiences, causing collecting information from neighbor nodes will also bring in noises.
\end{itemize}

%% file: 6_conclusions.tex
\section{Conclusion and Future Work}
In this work, we emphasize the importance of exploiting multiple interactions in CARS.
Towards this end, we first convert the features of users, items, and contexts into an attributed graph involving the contexts as edges between user and item nodes.
We then develop a new model, GCM, which captures the interactions among multiple user behaviors via graph neural networks, and then models the interactions among features of individual behavior via factorization machine.
To demonstrate the effectiveness of GCM, we test it on three public datasets, and it shows significant improvements over the state-of-the-art CF and CARS baselines.
Extensive experiments also are conducted to verify the rationality of attributed graph and offer insights into how the representations benefit from such graph learning.

Organizing user behaviors with contextual information in graphs is a promising direction to build an effective context-aware recommender.
It helps build strong representations for users and items.
However, GCM simply unifies all context features as an edge, neglecting dynamic characteristics of some contexts (\eg time) and hardly capturing dynamic preference of users~\cite{45}.
In future, we plan to build dynamic graphs based on contextual information, instead of one static graph, and devise a dynamic graph neural network.
Furthermore, rich side information is beneficial for explaining diverse intents behind user behaviors~\cite{46}.
We hence plan to model user-item relationships at a granular level of user intents to generate disentangled representations~\cite{47}.